\documentclass[10pt,leqno]{article}
\usepackage{graphicx}
\usepackage{indentfirst,csquotes}

\usepackage{amssymb,amsthm,amsmath}
\usepackage{xcolor,paralist,hyperref,titlesec,fancyhdr,booktabs,etoolbox}

\usepackage{microtype}
\usepackage{siunitx}
\usepackage{mhchem}
\usepackage{fancyvrb}

\newcommand*{\cpp}%
  {C\kern-0.05em\raisebox{0.18ex}{\scalebox{0.9}{+\kern-0.04em+}}}

\hypersetup{ colorlinks=true, linkcolor=black, filecolor=black, urlcolor=black }

\usepackage{authblk}

\begin{document}

\title{Optimization of Magnetized Electron Cooling with JSPEC} 
\date{\today}

\author[1]{Stephen J. Coleman\thanks{coleman@radiasoft.net}}
\author[1]{David L. Bruhwiler}
\author[1]{Dan T. Abell}
\author[1]{Boaz Nash}
\author[1]{Ilya~Pogorelov} 
\author[2]{He~Zhang}
\affil[1]{RadiaSoft, LLC, Boulder, CO}
\affil[2]{Thomas Jefferson National Accelerator Laboratory, Newport News, VA}

\maketitle


\begin{abstract}
The Electron-Ion-Collider (EIC) will be a next-generation facility located at Brookhaven National Laboratory (BNL), built with the goal of accelerating heavy ions up to 275~GeV.
To prevent ion beam size growth during the acceleration phase, cooling techniques will be required to keep the beam size from growing due to intra-beam scattering.
The JSPEC (JLab Simulation Package for Electron Cooling) \cpp\ package is a tool designed to numerically model magnetized and unmagnetized cooling through friction forces between co-propagating electron and ion bunches.
 Here we describe a feature that has been added to the JSPEC package, which implements a Nelder-Mead Simplex optimization algorithm to allow a user to optimize certain beam parameters in order to achieve a target cooling time. 
\end{abstract} 

\bigskip

\noindent 
\section{Introduction}
\begin{figure}[!t]
\centering
\includegraphics[width=3.in]{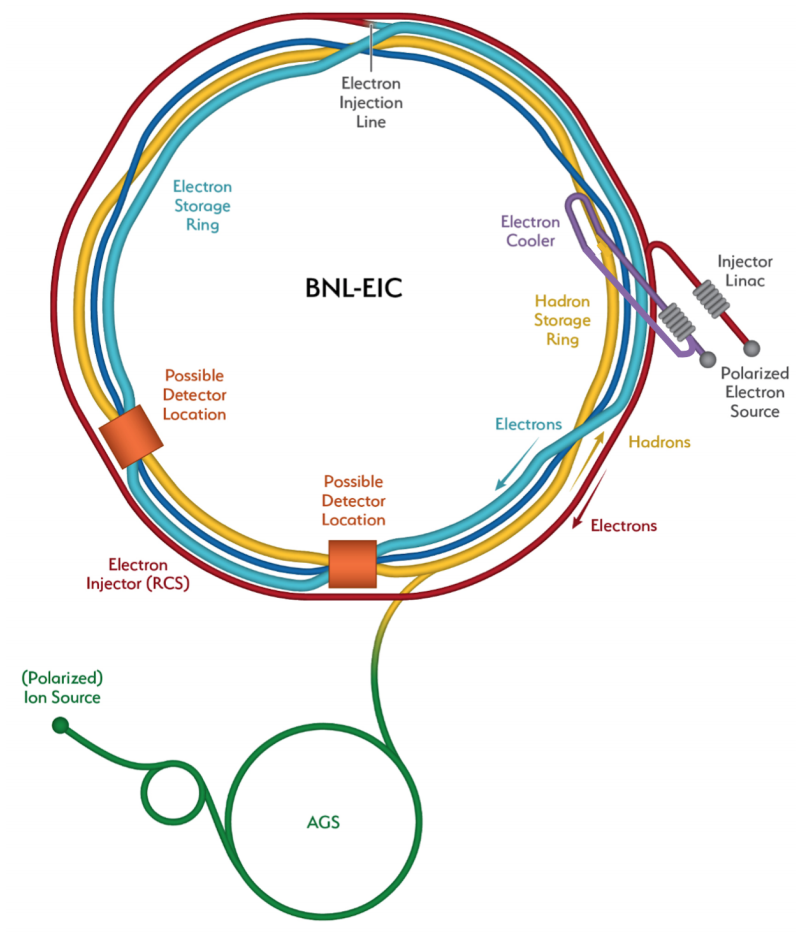}
\caption{Current overview of the electron-ion collider concept\cite{eRHIC_CDR}. The electron cooler is at 3 o'clock.}
\label{fig:eRHIC}
\end{figure}

The Electron-Ion-Collider (EIC), the layout of which is shown in figure~\ref{fig:eRHIC},  will be a next-generation facility located at Brookhaven National Laboratory (BNL), built with the goal of accelerating heavy ions up to \SI{275}{GeV} and luminosities up to \SI{e34}{cm^{-2}\second^{-1}}~\cite{eRHIC_CDR}.
To achieve these luminosities, ion beams will need to have high intensity and low momentum spread.
Ion beam size growth arises due to stochastic interactions associated with Intra-Beam Scattering (IBS). 
The growth can be slowed or balanced by introducing some form of cooling tuned to the characteristics of the beam.
Initial design studies plan for strong cooling of the ion beam using a coherent electron cooling technique, but the implementation of this technique depend on future R\&D efforts.
Magnetized electron cooling could be an alternative or a backup plan for controlling growth should the primary effort fall short \cite{eRHIC_CDR,LowEnergyCooling}.

In magnetized electron cooling, stochastic Coulomb collisions of individual ions in an ion bunch occur with co-propagating electrons that are confined in a solenoidal section. 
Interactions with electrons following `frozen' Larmor trajectories cause the ensemble of ions in the bunch to feel a drag, reducing the momentum spread through dynamical friction.
 As this friction force is reducing the average velocity of the ion beam in the beam rest frame, it can be considered as `cooling' the ion beam.
The possible benefits of magnetized electron cooling are reviewed in \cite{Parkhomchuk_1991, Larsson_1995, MESHKOV19971}.

\section{JSPEC}
The JLab Simulation Package for Electron Cooling (JSPEC) package is an open-source \cpp\ package originally developed at the Thomas Jefferson National Accelerator Facility~\cite{JSPEC1,JSPEC2}.
 In this package, friction force models and IBS models are applied to a model beam orbiting an accelerator with MAD-X lattice and other properties supplied by the user.
 JSPEC was extensively benchmarked against the Betacool electron cooling simulation package\,\cite{BETACOOL}, an older package based on MAD8 that also models IBS and cooling. 

The JSPEC dynamic calculation models the evolution of an ion bunch over time, after many passes in the cooler. 
The user has the option to produce dynamic simulations in one of two ways, either propagating with moments of distributions or with individual macro-particles.
In the former, the properties of representative ions are drawn randomly from the initial moments of each distribution. 
At the conclusion of each step in the dynamic calculation, new moments are calculated after perturbations from IBS and cooling kicks have been applied to the ensemble of representative ions. At the beginning of the subsequent step, new representative ions are initialized with properties drawn from those perturbed moments.  
In this way, the distributions are guaranteed to stay Gaussian. 
In the macro-particle method, initial properties of macro-particles are drawn from Gaussian distributions as before, but in the series of dynamic simulation steps each macro-particle experiences its own history of cooling and IBS kicks. 
The dynamics and the summary statistics are drawn from the ensemble of independent macro-particles. 
This is a more accurate representation of the physics, and non-Gaussian distributions are possible. For the first time step, where initial cooling rates are calculated and reported, these two methods yield identical results. 

\subsubsection{Rate Calculation}
JSPEC reports the rate of change of the emittance in horizontal, vertical, and longitudinal directions in units of 1/sec.
The initial rate of change is also broken down into components caused by IBS and by electron cooling. 
Initial IBS rates are calculated after initializing an ion bunch with known emittance values, propagating that bunch forward one step to induce IBS kicks (but not cooling kicks) to a set of macro-particles. 
The emittances are then re-calculated from the perturbed distributions. 
The IBS rate is simply the difference in the emittances divided by the simulation time step.
Two IBS models are available in JSPEC, the Bjorken-Mtingwa model~\cite{BM} and the Martini model~\cite{Martini}.
The initial cooling rate calculation is performed in a similar way to the IBS rate calculation. 
The rate is calculated by determining the difference in the emittance before and after cooling kicks are applied in the absence of IBS kicks, and dividing that difference by the time step.

\begin{figure}[!t]
\centering
\includegraphics[width=3.5in]{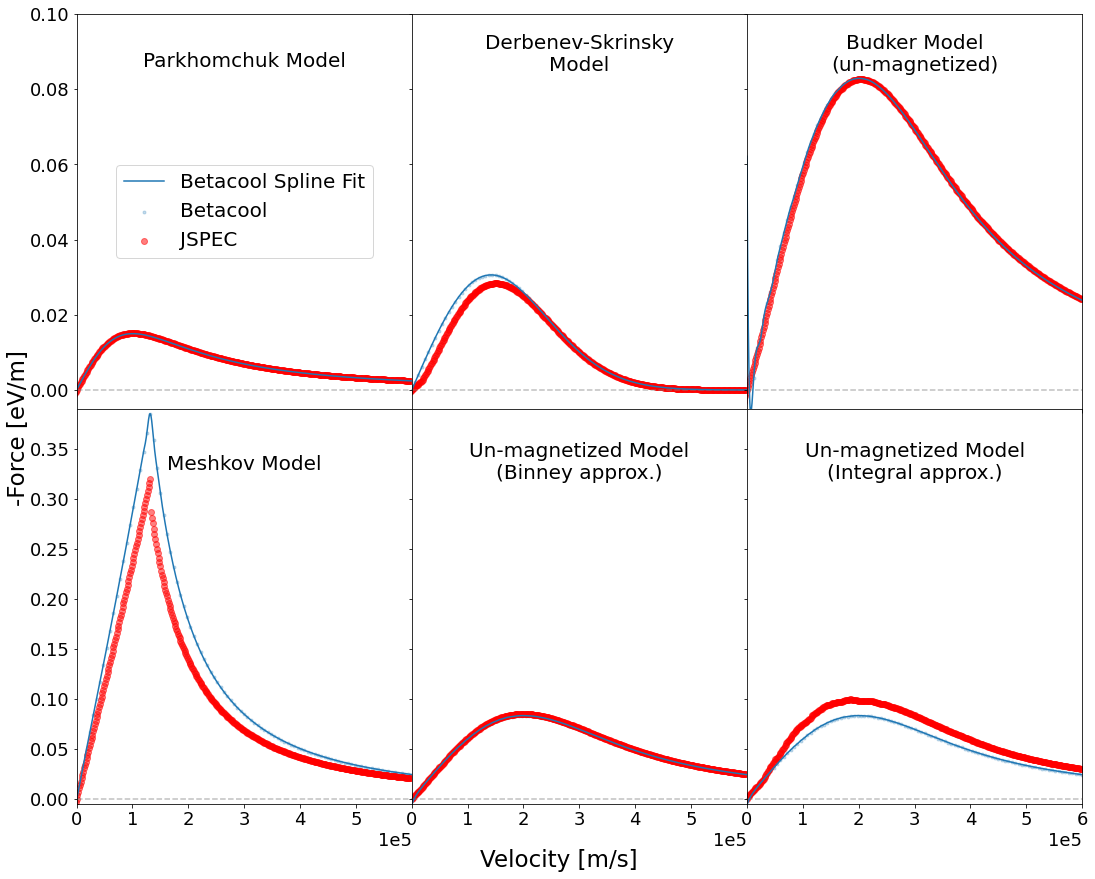}
\caption{Friction force curves as modeled in JSPEC and in Betacool.}
\label{fig:forces}
\end{figure}

Several new friction force models were added to the JSPEC package, bringing it in line with the set of models available in Betacool. 
This set includes the Derbenev \& Skrinsky model~\cite{DS}, the Meshkov asymptotic model~\cite{Meshkov}, the Budker un-magnetized model~\cite{Budker}, and two other numerical approximations of the first-principles unmagnetized model~\cite{BETACOOL}. 
Implementation of these models required numerical evaluation of indefinite integrals. 
These were solved through numerical routines provided by the GSL library~\cite{GSL}, and these calculations were parallelized through use of the OpenMP package~\cite{OpenMP}. These friction force formulas in JSPEC have been benchmarked with BETACOOL as shown in figure~\ref{fig:forces}.

\subsection{Optimization}
There are many parameters that may affect the cooling rate calculated for a proposed magnetized electron cooler. 
We have implemented a Nelder-Mead simplex optimization algorithm in JSPEC, which allows users to search a multi-dimensional parameter space for a set of values that meets their design needs. 
A full listing of parameters that may be varied within the optimization algorithm is shown in Table~\ref{tab:opt}.

\begin{table}
\centering
\begin{tabular}{lcc}
\toprule
  Parameter & Variable & \texttt{i} values\\
\midrule
  Ion Twiss $\alpha_{i}$  & \texttt{alpha\_i}     & \texttt{v,h} \\
  Ion Twiss $\beta_{i}$   & \texttt{beta\_i}      & \texttt{v,h} \\
  $e^-$ bunch RMS width $\sigma_{i}$
                          & \texttt{sigma\_i}     & \texttt{x,y,s} \\
  Dispersion              & \texttt{disp\_i}      & \texttt{v,h} \\
  Dispersion Derviative   & \texttt{disp\_der\_i} & \texttt{v,h} \\
  Cooler Magnetic Field $B$ & \texttt{bfield}     & --- \\
  $e^-$ Temperature       & \texttt{temp\_i}      & \texttt{tr,long} \\
  \# of electrons         & \texttt{n\_electron}  & --- \\
\bottomrule
\end{tabular}
\vspace{\baselineskip}
\caption{A full listing of variables that users may make available to the optimizer. Variables that are not initialized within the optimize section of the input file are held fixed. Here replace \texttt{i} by \texttt{v} or \texttt{h} for vertical or horizontal variables respectively.}
\label{tab:opt}
\end{table}

The Nelder-Mead simplex algorithm~\cite{NelderMead}, available within the GSL library~\cite{GSL}, was implemented because it is a gradient-free method, which makes it insensitive to the statistical noise encountered when cooling rates are repeatedly calculated with multiple independent simulations using a finite number of macro-particles. 
Each of the macro-particle parameters are Gaussian distributed and are independently drawn at each optimization step.
While the parameter space may be smooth in the limit of an infinite number of simulated macro-particles, the statistical noise associated with smaller numbers of macro-particles can lead to problems when evaluating the gradient.
The impact of this statistical noise is further mitigated by sampling a large number of macro particles, at the cost of longer simulation times (the default value is \num{e7}).

\subsection{Cost function}

The cost function for this optimization approach uses existing JSPEC cooling rate calculations.
The cooling rate is then converted into an approximate cooling time $T_i$ in a particular direction, $i \in x,y,s$. In units of minutes, this time is given by
\begin{equation}
  T_i = 60 \frac{1}{R_i} 
\end{equation}
where $R_i$ is the initial cooling rate in the $i$ direction. 
The value being minimized within the cost function $C$ is then
\begin{equation}
C_i = |T_0 - T_i|
\end{equation}
where the target cooling time is $T_0$. 
Note that cooling time and thus optimization can only be performed in one direction, so users must decide to prioritize cooling in a longitudinal or transverse direction based on their design constraints.
Without a defined target time, the cost function would effectively behave with $T_0=0$, leading the algorithm to select undesirable or extreme values for the free parameters, particularly for parameters that have approximately linear relationships with cooling time. 
In that case, the optimization algorithm finds reductions to the cost function by repeatedly varying a single parameter, ignoring all others.

\subsection{Uniqueness of results}

For an optimization problem with a large number of free parameters, there exist an infinite number of possible solutions along a locus in the multi-dimensional parameter space.
As a trivial example, consider an optimization problem with three free parameters, with selected values that satisfy the optimization condition at $\alpha=\alpha_0$, $\beta=\beta_0$, and $\gamma=\gamma_0$.
Now assume the two parameters $\alpha$ and $\beta$ are anti-correlated, meaning that an infinitesimal increase in one parameter coupled with an infinitesimal decrease in the other will yield the same result.
While this may remain true for perceptible deviations, a significant change (say, $\alpha \rightarrow \alpha_1$ and $\beta \rightarrow \beta_1$) may cause the third free parameter to compensate as $\gamma\rightarrow\gamma_1$.
Now, at this new point, infinitesimal changes about $\alpha_1$ can be compensated with anti-correlated changes about $\beta_1$.
Thus, many equivalent solutions exist and repeated attempts at optimization with identical initial conditions may not yield the same sets of optimal parameters.

\subsection{Searching beyond local minima}

Nelder-Mead Simplex optimization procedures may sometimes fall into local minima and fail to emerge.
For this reason, the optimization procedure is attempted many times, each with slightly different initial conditions. 
The initial conditions for each attempt are drawn randomly from a Gaussian distribution, using the user-provided starting point as the mean of the distribution. 
For each attempt, initial conditions are then drawn at random and used to initialize the Nelder-Mead Simplex algorithm. 
After a number of attempts (15 by default) the optimal set of parameters producing the minimum value of the cost function $C$ from any of these attempts is stored in a plaintext output file, \texttt{best.txt}.

If the Nelder-Mead Simplex suggests a parameter that is unphysical (e.g.~a negative bunch length, or a negative electron density), the optimization routine will alert the user that an unphysical attempt has been made and will discard the suggestion.

\subsection{Parameter Scans}

\begin{figure}[!t]
\centering
\includegraphics[width=3.5in]{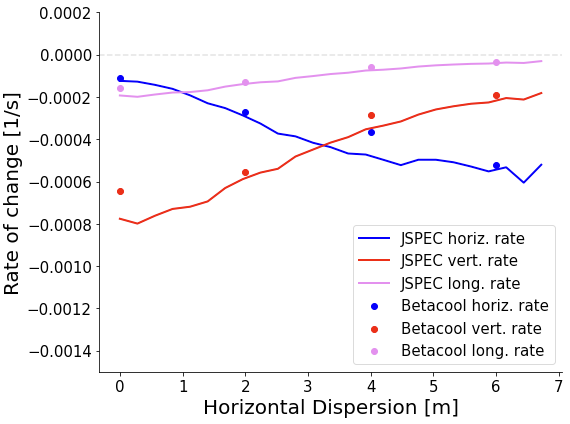}
\caption{Parameter scan of the horizontal dispersion parameter performed in JSPEC compared to single-point sampled values in Betacool.}
\label{fig:scan}
\end{figure}

Users may hold $N-1$ parameter values constant and scan values of the holdout parameter in order to see its effect on cooling rates.  Betacool does not have a built-in support for parameter scans.
A comparison of scanned values for matching configurations in JSPEC and in Betacool is shown in Figure~\ref{fig:scan}.

A reasonable strategy for optimization would be to allow the optimization algorithm to find a suitable set of parameters, then examine the 1-d parameter scans of each free parameter in order to understand the nature of the dependence of cooling time on each parameter in the local region of parameter space.
This can also be useful for practical purposes.
For example, an optimum set of parameters may include a solenoid magnetic field of $>\SI{10}{T}$, a difficult and expensive element to produce. 
A parameter scan might reveal that a lower solenoid field strength will suffice to satisfy the design objectives of the cooler.
The syntax for configuring a JSPEC parameter scan are shown in Appendix B.

\section{Examples for EIC}
The EIC will accelerate ions, from protons ($Z=1$) to \ce{Au+} ($Z=79$) to collision energies ranging from 41--275\,GeV. 
We will construct examples at both ends of the ion mass spectrum and use them to demonstrate the optimization and parameter scan features of JSPEC. 
We will also construct an example to demonstrate magnetized electron cooling at pre-injection. 
We will select the Parkhomchuk~\cite{Parkhomchuk} friction force model for all cases. 

\subsection{Cooling at pre-injection}
One concept being considered for cooling at EIC is to cool protons at pre-injection, with ion beam energies of 23.8\,GeV. 
This is favorable for magnetized electron cooling, because the friction force scales with $1/\gamma^2$. 
Cooling at lower energies is generally easier. 
The parameters being considered for this concept are shown in Table~\ref{tab:pre-injection}. 

\begin{table}
\centering
\begin{tabular}{lc}
\toprule
  Parameter Name & value  \\
\midrule
  Species                       & Proton \\
  Ion Energy [GeV]              & 23.8 \\
  Bunch Intensity [$10^{10}$]   & 2.6 \\
  Beam Current [A]              & 0.69 \\
  $\beta$ horizontal [cm]       & 150-200 \\
  $\beta$ vertical [cm]         & 250-200 \\
  $\epsilon_{\mathrm{Horiz}}$ (Norm.) [$\mu$m] & 2.7 \\
  $\epsilon_{\mathrm{Vert}}$  (Norm.) [$\mu$m] & 0.25 \\
  $\Delta p / p $ [$10^{-4}$]   & 10.3 \\
  Bunch Length [cm]             & 60 \\
  Length of cooling section [m] & 130 \\
\bottomrule
\vspace{\baselineskip}
\end{tabular}
\caption{Table of low-energy cooling parameters.
         Values from \cite{LowEnergyCooling}.}
\label{tab:pre-injection}
\end{table}

\subsubsection{Preliminary test}
We start the optimization procedure by allowing only a few parameters to float: the transverse beam sizes $\sigma_x$ and $\sigma_y$, the solenoid field strength $B$, and the number of electrons in a bunch. 
We set the target cooling time to $T_0=20$\,minutes. The initialization code for this is shown in Appendix A and the results are listed in Table~\ref{tab:EIC_ini_opt}.
We see that the optimization has yielded a cooling time within a second of 20 minutes. 
While this satisfies the target time, we notice that the electron density is quite high from an operational standpoint. 

\begin{table}
\centering
\begin{tabular}{lc}
\toprule
  Parameter Name &  optimization result \\
\midrule
  $\sigma _x$ [$\mu$ m]          & 48.67 \\
  $\sigma _y$ [$\mu$ m]          & 317.94 \\
  B field [T]    & 1.93 \\
  $N_{e}$ [$10^{10}$]       & 0.94 \\
  Cost function $C$ [min] & \num{4.29e-4} \\
\bottomrule
\end{tabular}
\vspace{\baselineskip}
\caption{Table of optimized parameters after the preliminary optimization for 20 minute cooling times  for beam conditions shown
         in Table~\ref{tab:pre-injection}.}
\label{tab:EIC_ini_opt}
\end{table}


\subsubsection{Alternate optimizer parameterization}
\begin{figure}[!t]
\centering
\includegraphics[width=3.6in]{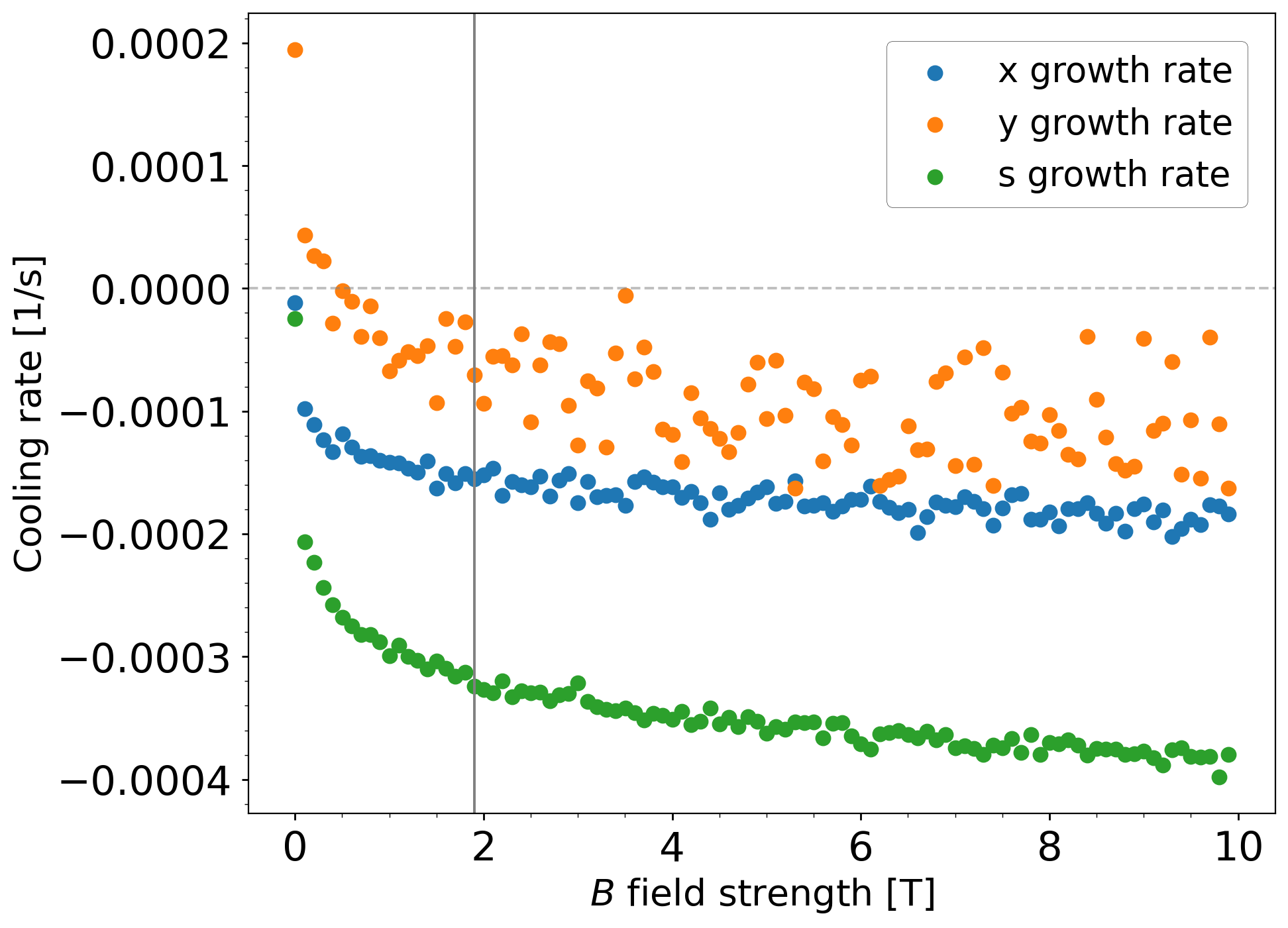}
\caption{Parameter scan of the electron bunch length parameter around the central value determined from the optimization algorithm. 
The grey vertical line marks the value suggested by the optimizer, while the dashed horizontal line denotes the boundary between ion beam heating (positive values) or cooling (negative) values.}
\label{fig:scan_bfield}
\end{figure}

Examining the 1-dimensional dependence of the cooling time on a particular parameter may inform excursions away from optimum values based on such practicalities. 
While the friction force models may inform the dependence on some parameters, for example a linear dependence on $n_e$, non-linear relationships may be observed with sampling in a parameter scan. 
In Figure~\ref{fig:scan_bfield} increasing $B$-field values, holding all other parameters fixed, yields shorter cooling times, but the relationship is asymptotic. 
A user may judge whether a 2\,Tesla solenoid is sufficient for the cooling goals. 

We can now fix the B-field value and run the optimization algorithm again, with a different set of floating parameters. 
The free parameters were the electron bunch length $\sigma_s$, the number of electrons in the bunch $n_e$, the transverse electron temperature $T_{\perp}$, the longitudinal electron temperature $T_{\parallel}$, and the $\beta_v$ and $\beta_h$ of the cooler.
 The target was a 20 minute cooling time in the longitudinal direction, and then separately in the vertical direction, to match the EIC magnetized electron cooling design study~\cite{LowEnergyCooling}. 
 Results are summarized in Table~\ref{tab:EIC_lowE_opt}.

\begin{table}
\centering
\begin{tabular}{lcc}
\toprule
  Parameter Name & $s$ optimization & $y$ optimization \\
\midrule
  $\beta_y$ [m]           & 99.8     & 100.1 \\
  $\beta_y$ [m]           & 98.8     & 102.0 \\
  $N_{e}$ [$10^{10}$]     & 1.06     & 1.04 \\
  $T_{\perp}$ [eV]        & 0.014    & 0.011 \\
  $T_{\parallel}$ [eV]    & 0.01     & 0.013 \\
  $\sigma_{s}$ [cm]       & 4.48     & 4.90 \\[3pt]
  Cost function $C$ [min] & \num{1.12e-4} & \num{1.96e-4} \\
\bottomrule
\end{tabular}
\vspace{\baselineskip}
\caption{Table of optimized parameters for 20 minute cooling times
         in the longitudinal direction and the vertical direction for beam conditions shown
         in Table~\ref{tab:pre-injection}.}
\label{tab:EIC_lowE_opt}
\end{table}


This satisfies the target cooling time with the inverse of the initial rate, but it's worth looking at the long-term dynamic behavior of an ion bunch that passes through the cooler on multiple orbits. An approximate bunch of macro-particles representing the whole ion bunch may be simulated with JSPEC. 
Each of the macro-particles evolves independently, and summary statistics and emittance are then calculated from the ensemble.

\subsection{Cooling for EIC}
We can use the optimization algorithm to explore the use of magnetized electron cooling within the EIC hadron storage ring. 
Here we simulate the lowest operating energy range within the full range of ion masses. 
The optimization algorithm was initialized with the fixed values shown in Table~\ref{tab:EIC}, with either protons or gold ions. 
The cooling target time was set to be 20 minutes in the longitudinal direction in both cases. The optimized parameters are presented in Table~\ref{tab:EIC_opt}.

\begin{table}
\centering
\begin{tabular}{lcc}
\toprule
  Parameter Name              & proton & \ce{Au+} \\
\midrule
  Ion Energy [GeV]            & 41     & 41  \\
  Bunch Intensity [\num{e10}] & 2.6    & 0.036 \\
  Beam Current [A]            & 0.69   & 0.41 \\
  $\beta$ horizontal [cm]     & 90     & 90 \\
  $\beta$ vertical [cm]       & 7.1    & 4 \\
  $\epsilon_\text{Horiz}$ (Norm.) [$\mu$m] & 2.7  & 3.0 \\
  $\epsilon_\text{Vert}$ (Norm.)  [$\mu$m] & 0.25 & 0.3 \\
  $\Delta p / p$ [\num{e-4}]  & 10.3   & 10.0 \\
  Bunch Length [cm]           & 7.5    & 11.6 \\
\bottomrule
\end{tabular}
\vspace{\baselineskip}
\caption{Table of EIC parameters. Proton and gold ion values are taken
         from Table~3.3 and Table~3.5 in~\cite{eRHIC_CDR} respectively.}
\label{tab:EIC}
\end{table}

\begin{table}
\centering
\begin{tabular}{lcc}
\toprule
  Parameter Name & proton & \ce{Au+} \\
\midrule
  Ion Energy [GeV] & 41 & 41  \\
  $B$-field [T] & 2.07 & 2.35 \\
  $N_{e}$ [$10^{10}$] & 1.22 & 0.156 \\
  $T_{\perp}$ [eV] & 0.011 & 0.013 \\
  $T_{\parallel}$ [eV] & 0.009 &  0.008 \\
  $\sigma_{s}$ [cm] & 5.8 & 5.8 \\[3pt]
  Cost function $C$ [min] & $4.8\times 16^{-5}$ & $1.26\times 16^{-4}$ \\
\bottomrule
\vspace{\baselineskip}
\end{tabular}
\caption{Table of optimized parameters for 20 minute cooling times
         in the longitudinal direction for beam conditions shown
         in Table~\ref{tab:EIC}.}
\label{tab:EIC_opt}
\end{table}

\section{Conclusion}
We have shown techniques that can be used to optimize the design of an arbitrary magnetized electron cooler. 
A Nelder-Mead Simplex optimization algorithm was introduced to the JSPEC magnetized electron cooling simulation code in order to sample many possible cooler configurations. 
We then demonstrated these techniques with examples relevant to the proposed electron ion collider. 
The optimized sets of parameters were then validated through dynamic simulations, also generated with JSPEC. 
Readers may simulate the optimized sets of parameters for themselves using the cloud-based Sirepo interface for JSPEC available at \url{https://sirepo.com/jspec}.

$\,$
\clearpage
\appendix
\section{Code for defining a parameter scan}
Much of the previous appendix conveys for a parameter scan. 
The changes which initialize a parameter scan, in this case for the magnetic field strength $B$ (\texttt{bfield}), are shown here:

\begin{Verbatim}[frame=single]
...
section_optimization
        bfield = 1.0e-4
        bfield = 10.0
        steps = 100
section_run
        total_expansion_rate
        optimize_cooling
\end{Verbatim}
Here the repeated definition of the variable initializes the parameter scan and activates the \texttt{steps} variable, which defines the granularity of the scan.
 The output is stored in a plaintext \texttt{scan.txt} file showing the cooling rate for $x,y$, and $s$ directions at each step. 
 
\section{Input code for pre-injection cooling optimization}
The full plaintext input file parsed by JSPEC is shown below, for an optimization problem. 
This input file requires the lattice file to be in Mad-X Twiss parameter format (\texttt{.tfs} file type), and that the file be present in the working directory. 
Values defined in \texttt{section\_optimization} over-ride values for the same parameter defined in earlier sections.  
All parameters that are not defined in the \texttt{section\_optimization} are fixed, and are not allowed to vary with the optimization algorithm.
\begin{Verbatim}[frame=single]
section_scratch
        ion_mass = 938.272
        ion_ke = 23800.0
        ion_gamma = 1 + ion_ke/ion_mass
section_ion
        charge_number = 1
        mass = ion_mass
        kinetic_energy = ion_ke
        norm_emit_x = 2.5e-06
        norm_emit_y = 2.5e-06
        momentum_spread = 0.001
        particle_number = 2.6e10
        rms_bunch_length = 0.6
section_ring
        lattice = Lattice.tfs
section_ibs
        nu = 100
        nv = 100
        log_c = 20.6
        coupling = 0.0
section_cooler
        length = 130.0
        section_number = 1
        magnetic_field = 5.06
        bet_x = 150
        bet_y = 150
        disp_x = 0.0
        disp_y = 0.0
        alpha_x = 0.0
        alpha_y = 0.0
        disp_dx = 0.0
        disp_dy = 0.0
section_e_beam
        gamma = ion_gamma
        tmp_tr = 0.0001
        tmp_l = 0.01
        shape = bunched_gaussian
        radius = 0.009
        current = 4.0
        sigma_x = 0.0002
        sigma_y = 0.0002
        sigma_z = 0.07
        length = 0.05
        e_number = 5e10
section_ecool
        sample_number = 100000.0
        ion_sample = MONTE_CARLO
        force_formula = PARKHOMCHUK
section_run
        create_ion_beam
        create_ring
        create_e_beam
        create_cooler
section_simulation
        ibs = on
        e_cool = on
        time = 2000.0
        step_number = 100
        output_file = JSPECdump.SDDS
        model = RMS
        ref_bet_x = 10.0
        ref_bet_y = 10.0
        ref_alf_x = 0.0
        ref_alf_y = 0.0
        ref_disp_x = 0.0
        ref_disp_y = 0.0
        ref_disp_dx = 0.0
        ref_disp_dy = 0.0
section_optimization
        sigma_x = 2e-4
        sigma_y = 2e-4
        bfield = 2.0
        n_electron = 1.5
        axis = s
        time = 20
section_run
        total_expansion_rate
        optimize_cooling
\end{Verbatim}
The best fit parameters will be stored in a plaintext file in the working directory called \texttt{bests.txt}.

$\,$

\bibliographystyle{unsrt}
\bibliography{optimization}

\begin{thebibliography}{10}

\bibitem{eRHIC_CDR}
Ferdinand Willeke and J.~Beebe-Wang.
\newblock Electron ion collider conceptual design report 2021.
\newblock 2 2021.

\bibitem{LowEnergyCooling}
A.~Fedotov et~al.
\newblock Low energy cooling for electron ion collider.
\newblock Technical Report BNL-220686-2020-TECH, Brookhaven National Laboratory, 12 2020.

\bibitem{Parkhomchuk_1991}
V~V Parkhomchuk and A~N Skrinsky.
\newblock Electron cooling: physics and prospective applications.
\newblock {\em Reports on Progress in Physics}, 54(7):919--947, jul 1991.

\bibitem{Larsson_1995}
M~Larsson.
\newblock Atomic and molecular physics with ion storage rings.
\newblock {\em Reports on Progress in Physics}, 58(10):1267--1319, oct 1995.

\bibitem{MESHKOV19971}
I.N. Meshkov.
\newblock Electron cooling — the first 30 years and thereafter.
\newblock {\em Nuclear Instruments and Methods in Physics Research Section A: Accelerators, Spectrometers, Detectors and Associated Equipment}, 391(1):1--11, 1997.

\bibitem{JSPEC1}
{JSPEC} source code.

\bibitem{JSPEC2}
H.~Zhang et~al.
\newblock The latest code development progress of {JSPEC}.
\newblock In {\em Proc. of NAPAC2019}, pages 539--541, Lansing, MI, 2019.

\bibitem{BETACOOL}
I.~Meshkov et~al.
\newblock {\em {BETACOOL} Physics Guide}.
\newblock 2007.

\bibitem{BM}
James~D. Bjorken and Sekazi~K. Mtingwa.
\newblock {Intrabeam Scattering}.
\newblock {\em Part. Accel.}, 13:115--143, 1983.

\bibitem{Martini}
M.~Martini.
\newblock Intrabeam scattering in the {ACOOL-AA} machines.
\newblock Technical Report CERN PS/84-9 AA, CERN, May 1984.

\bibitem{DS}
Ya.S. Derbenev and A.N. Skrinsky.
\newblock {\em Fyzika plasmy}, 4:492, 1978.

\bibitem{Meshkov}
I.~N. Meshkov.
\newblock {Electron cooling: Status and perspectives}.
\newblock {\em Phys. Part. Nucl.}, 25:631--661, 1994.

\bibitem{Budker}
G.~Budker.
\newblock In {\em Proc. Intern. Symp. Electron and Positron Storage Rings}, pages 11--1--1, Saclay, 1966.

\bibitem{GSL}
M.~Galassi et~al.
\newblock {\em {GNU} Scientific Library Reference Manual (3rd Ed.)}.
\newblock Network Theory Ltd., 2009.

\bibitem{OpenMP}
Leonardo Dagum and Ramesh Menon.
\newblock {Open{MP}}: An industry-standard {API} for shared-memory programming.
\newblock {\em {IEEE} Comput. Sci. Eng.}, 5:46–55, January 1998.

\bibitem{NelderMead}
J.A. Nelder and R.~Mead.
\newblock A simplex method for function minimization.
\newblock {\em Computer Journal}, 7:308--313, 1965.

\bibitem{Parkhomchuk}
V.~V. Parkhomchuk.
\newblock {New insights in the theory of electron cooling}.
\newblock {\em Nucl. Instrum. Meth. A}, 441:9--17, 2000.

\end{thebibliography}

\end{document}